\documentclass{rnaastex}
\usepackage{amsmath, amssymb, bm}

\begin{document}

\title{K2SUPERSTAMP: The release of calibrated mosaics for the {\em Kepler/K2} Mission}

\correspondingauthor{Ann Marie Cody}
\email{annmarie.cody@nasa.gov}

\author{Ann Marie Cody}
\affiliation{NASA Ames Research Center, Moffett Blvd, Mountain View, CA 94035, USA}
\affiliation{Bay Area Environmental Research Institute, 625 2nd St Ste. 209, Petaluma, CA 94952}

\author{Geert Barentsen}
\affiliation{NASA Ames Research Center, Moffett Blvd, Mountain View, CA 94035, USA}
\affiliation{Bay Area Environmental Research Institute, 625 2nd St Ste. 209, Petaluma, CA 94952}

\author{Christina Hedges}
\affiliation{NASA Ames Research Center, Moffett Blvd, Mountain View, CA 94035, USA}
\affiliation{Bay Area Environmental Research Institute, 625 2nd St Ste. 209, Petaluma, CA 94952}

\author{Michael Gully-Santiago}
\affiliation{NASA Ames Research Center, Moffett Blvd, Mountain View, CA 94035, USA}
\affiliation{Bay Area Environmental Research Institute, 625 2nd St Ste. 209, Petaluma, CA 94952}

\author{Jos\'e Vin\'icius de Miranda Cardoso}
\affiliation{NASA Ames Research Center, Moffett Blvd, Mountain View, CA 94035, USA}
\affiliation{Bay Area Environmental Research Institute, 625 2nd St Ste. 209, Petaluma, CA 94952}


\section*{{\em K2} Mission Superstamp Observations}

In operation since 2014, the NASA {\em K2} Mission \citep{howell} is using the {\em Kepler} telescope to photometrically monitor tens of thousands 
of stars and other astronomical objects over the course of 70--80 day campaigns. Observations are focused on regions in the ecliptic plane. The 
mission publicly releases all of its imaging data as well as light curves for the majority of targets.

With a $\sim 10\arcdeg\times 10$\arcdeg\ field of view, {\em Kepler}'s full frame images are too large to efficiently downlink more than twice per 
campaign. Therefore, images of most targets are provided as small postage stamp cut-outs, of order $10\times 10$ pixels. For uncrowded point 
sources, precise differential photometry can be extracted from these stamps using simple aperture photometry (with, e.g., \texttt{pyke}; PyKE 
Contributors 2017). Light curve production becomes more problematic for extended sources such as open clusters, galaxies, and globular clusters. 
These spatially large targets are typically observed as ``superstamps,'' which consist of many smaller stamps that can be tiled together to form a 
complete image of the region of interest. While it is useful for photometry and data analysis to have the full superstamp images, to date the 
mission has only released the postage stamps-- leaving it up to the user to reassemble each superstamp. In this note, we detail a new public 
release of full {\em K2} superstamp images at 30 minute cadence for the four open clusters M35, M67, Ruprecht~147, and NGC~6530 (Lagoon Nebula 
Cluster, hereafter ``the Lagoon'').

\section*{Creation of FITS Images}

The {\em K2} pipeline uniformly processes all imaging data in each campaign. After basic calibrations, the postage stamps from all observing epochs 
in a given {\em K2} campaign are stacked into a target pixel file (``TPF'') and released through the Barbara A. Mikulski Archive for Space 
Telescopes (MAST)\footnote{https://archive.stsci.edu/k2}. This is the case regardless of whether the target is a single star or an extended object. 
Superstamp images thus consist of many target pixel files that must be unstacked in the time domain and stitched together in the spatial domain. We 
have created a pipeline to do just this, and we detail the process here. Python code is available on 
Github\footnote{https://github.com/amcody/SuperstampFITS}.

For every observing epoch (``cadence''), we seamlessly stitch the TPF frames together by using the detector coordinates. For superstamps that cross 
detector channel boundaries (i.e., M67 and the Lagoon), some of the stamps must be flipped about one axis before stitching. This is a result of the 
way detector coordinates are assigned; the readout direction reverses from one CCD channel to the next.

Final images consist of the 'FLUX' data from the original TPFs. As of this writing, the 'FLUX' data for M67 and Ruprecht~147 have been background 
subtracted by the {\em K2} pipeline, while for M35 and the Lagoon they have not. The final images are written to a FITS file. Header information 
from the original TPFs is carried forward. TPF header keywords intended for single targets are removed. We calculate and add `DATE-OBS' and 
`DATE-END' in UTC.

\begin{figure}[h]
    \centering
    \plotone{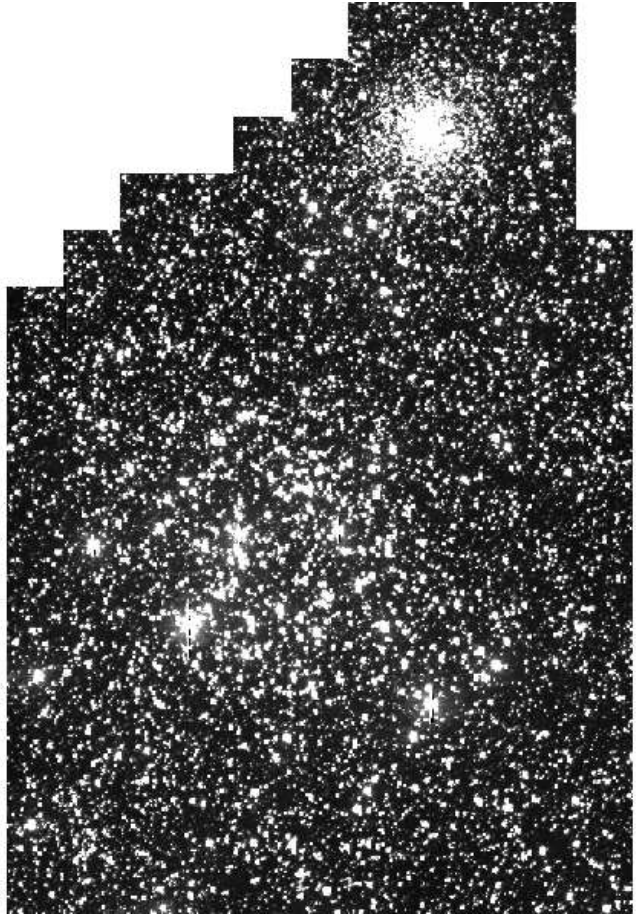}
    \caption{Example {\em K2} superstamp image of M35, which also includes the older NGC~2158 cluster toward the top.} 
    \label{m35image}
\end{figure}

The initial superstamp images that we generate do not have accurate world coordinate systems (WCS). The WCS keywords in the original TPFs apply to 
all epochs, despite the fact that there is pixel-level jitter of targets. We carried out a new astrometric calibration. We employed the 
Astrometry.net program \citep{lang}, which used triangles of USNO-B catalog stars to determine and apply a WCS solution. The user can accurately 
convert between right ascension and declination of an object and its time-dependent pixel position using this solution.

For most of the superstamp images, the Astrometry.net WCS solution is sufficient to produce an accurate translation between RA/Dec and $x$/$y$ 
pixel coordinates. But in the case of the Lagoon images, we found that nebulosity in the region prevented enough USNO-B stars from being detected 
for a good WCS solution. The WCS put in place by Astrometry.net displays a uniform shift of 1--2 pixels that is evident when overlaying 2MASS 
catalog positions on the images. To correct this problem, we selected 24 moderately bright 2MASS stars to recalibrate the WCS. For this task, we 
employed the {\em IRAF} \citep{tody} task \texttt{CCMAP} to solve for a new solution based on the known 2MASS coordinates and $x$, $y$ centroid 
positions determined by the PyKE tool \texttt{kepprf}. This new solution was then used to overwrite the previous FITS header keywords.

\section*{Public Superstamp Image Release}
The assembled and astrometrically calibrated superstamp FITS images for M35, M67, Ruprecht 147, and NGC 6530 are now available as a High Level 
Science Product on MAST\footnote{https://archive.stsci.edu/prepds/k2superstamp} with DOI 
\dataset[10.17909/T9M09M]{http://dx.doi.org/10.17909/T9M09M}. We provide a sample image of the M35 field in Figure~1. It is our hope that these new 
public products will spur further analyses of variability and planet searches among the open clusters observed by {\em K2}.

\acknowledgments This work acknowledges the use of the following computational
tools: \texttt{numpy}~\citep{numpy},
\texttt{astropy}~\citep{astropy}, and \texttt{pyke}~\citep{pyke}.


\begin{thebibliography}{}
\bibitem[AstroPy Project (2017)]{astropy} AstroPy Project, Submitted to ApJ, 2017.
\bibitem[Howell (2014)]{howell} Howell, S. B. 2014, PASP, 126, 398
\bibitem[Lang et al.\ (2010)]{lang} Lang D., Hogg D. W., Mierle K., Blanton M. and Roweis S. 2010, AJ, 139, 1782
\bibitem[PyKE Contributors (2017)]{pyke} PyKE Contributors, PyKE: Kepler, $\mathcal{K}\mathit{2}$ \& TESS Data Analysis Tools, 2017,
\url{http://doi.org/10.5281/zenodo.835583}.
\bibitem[Tody (1993)]{tody} Tody, D. 1993, "IRAF in the Nineties" in Astronomical Data Analysis Software and Systems II, A.S.P. Conference Ser., Vol 52, eds. R.J. Hanisch, R.J.V. Brissenden, \& J. Barnes, 173.  
\bibitem[van der Walt et al. (2011)]{numpy} van der Walt S., Colbert S.~C., Varoquaux G., The NumPy Array: a structure for efficient numerical computation, 2011, Computing in Science \& Engineering, 13, 22--30. 
\end{thebibliography}
\end{document}